# Quantify Change of Inertia and Its Distribution in High Renewable Power Grids Using PMU

Shutang You

*Abstract*— This paper proposed an approach to identify the change of inertia distribution in high renewable power systems. Using the footprints of electromechanical wave propagation at the distribution level, this approach provides a new and non-invasive way to aware the system inertia distribution for primary frequency response. Actual measurements and high renewable dynamic models validated effectiveness of the approach.

## I. Introduction

As the wind and solar photovoltaic (PV) outputs are highly variable, unit commitment becomes more uncertain and dynamic, while the requirements on system inertia increase in order to support primary frequency response. The reliability of the system is influenced by many factors including inertia. Therefore, assessing the inertia adequacy and its distribution has become a necessity but also a difficulty, especially in deregulated market environments. The conventional approach requires the information on event magnitude, for example the output of a given tripped generator. However, the event information is not usually instantly available by all utilities and ISOs. In addition, understanding how much inertia each region is contributing for primary frequency response is an important track information for regulatory organizations such as the North American Electric Reliability Corporation. This requirement can not be satisfied by a single estimation result on the system total inertia provided by the conventional method.

This paper presents a non-invasive approach to assess the system inertia distribution using frequency transient measurements from distribution level PMUs. In estimating the inertia distribution, this approach leverages the non-uniformity of the electromechanical wave propagation speed during system events that cause frequency deviations, which typically occur multiple times daily for large power systems. The proposed approach was validated by two datasets: the actual measurements from a wide-area synchronous measurement system at the distribution-level — FNET/GridEye and the simulation data of a high renewable model representing the future U.S. Eastern Interconnection (EI) transmission system.

## II. Non-Invasive Identification of Inertia Distribution Change

Electromechanical wave propagation is a transient process when the electrical angular frequency disturbance propagates from the event location to the rest of the system. The propagation speed is influenced by many factors, including the system topology, line parameters, voltage level, generator and load inertia.

Due to its complexity in nature, the continuum model was proposed to study electromechanical wave propagation theoretically [1]. In this ideal model, which considers all impact factors in a simple way, the square of the electromechanical wave propagation speed is inversely proportional to the inertia distribution: $|\vec{v}|^2 = \omega V^2 sin\theta/2|z|h$, where $v$ is the electromechanical wave propagation speed. $\omega$ is the electrical angular frequency (p.u.). $V$ is the voltage magnitude (p.u.). $\theta$ is the line impedance angle ($\theta \approx \pi/2$ in transmission networks). $|z|$ is the line impedance (p.u.). $h$ is the inertia of per unit length of the continuum model. This formula indicates that under high renewable penetration, the dominant impact factor of the propagation speed that varies on a daily basis will be the generation and load inertia distribution. Without losing generality, the relation between the propagation speed and inertia distribution is expressed as: $\vec{v} = f(h)$. In function $f$, $|\vec{v}|$ decreases monotonically with the increase of $h$. Thus, $h$ can be expressed as a function of $|\vec{v}|$:

$$h = \bar{f}^{-1}(|\vec{v}|) \qquad (1)$$

The electromechanical wave propagation speed $|\vec{v}|$ can be calculated based on the distribution level PMU measurement. The time delay of arrival (TDOA) of the frequency disturbance at one PMU differs from that at another PMU due to the difference in distance between the PMUs and the disturbance location. Thus the TDOA can be obtained by setting a threshold of frequency and recording the time when the frequency crosses the threshold. The propagation speed of each location could be obtained from its local footprint of wave propagation:

$$|\vec{v}| = 1 / \left| \frac{d\text{TDOA}}{d\vec{s}} \right| \qquad (2)$$

where $\vec{s}$ is the per unit distance along the direction of wave propagation. $d\text{TDOA}/d\vec{s}$ can be decomposed in to the longitude and the latitude direction as:

$$\begin{aligned}\frac{d\text{TDOA}}{d\vec{s}} &= \left| \frac{\partial \text{TDOA}}{\partial \overrightarrow{c_{lon}}} \right| \vec{e}_{lon} + \left| \frac{\partial \text{TDOA}}{\partial \overrightarrow{c_{lat}}} \right| \vec{e}_{lat} \\ &= \left| \frac{\partial \text{TDOA}}{|\overrightarrow{c_{lon}}| \partial \vec{e}_{lon}} \right| \vec{e}_{lon} + \left| \frac{\partial \text{TDOA}}{|\overrightarrow{c_{lat}}| \partial \vec{e}_{lat}} \right| \vec{e}_{lat}\end{aligned} \qquad (3)$$

where $|\overrightarrow{c_{lon}}|$ and $|\overrightarrow{c_{lat}}|$ are the coefficients of per unit distance at this location for one degree of longitude and latitude, respectively. Their values can be obtained from the Haversine formula [2].

Substitute (3) to (2), one can obtain:

$$h = \bar{f}^{-1}\left(1/\sqrt{\left|\frac{1}{|\vec{c}_{lon}|}\left|\frac{\partial \text{TDOA}}{\partial \vec{e}_{lon}}\right|\right|^2 + \left|\frac{1}{|\vec{c}_{lat}|}\left|\frac{\partial \text{TDOA}}{\partial \vec{e}_{lat}}\right|\right|^2}\right) \quad (4)$$

where $|\partial \text{TDOA}/\partial \vec{e}_{lon}|$ and $|\partial \text{TDOA}/\partial \vec{e}_{lat}|$ are the gradient values of the DTOA in the longitude and latitude direction for one degree, respectively.

Since TDOA is required at each point that needs to perform inertia estimation, PMUs may be inadequate in number or very unevenly distributed. Here, the two-dimension Biharmonic spline interpolation method [3] is applied to reconstruct TDOA at locations without PMU coverage based on the available measurements, which are scattered in $N$ locations:

$$\text{TDOA}_{x,y} = \nabla^4 \text{TDOA}(x,y) = \sum_{j=1}^{N} \alpha_j \delta\left((x,y) - (x_j, y_j)\right) \quad (5)$$

where $x$ and $y$ are the longitude and latitude coordinates. $(x_j, y_j)$ $j = 1, \ldots, N$ are coordinates of PMU-measured locations.

Furthermore, PMUs may have data quality and timing issues, such as leap second, year rollover, and loss of GPS signal. These factors may impact the inertia estimation through providing wrong TDOA. For robustness, the approach applied linear regression and the estimated location of the event to check the measured TDOA. The validation includes the following steps:

1) In the interpolation result $\text{TDOA}_{x,y}$, the point $(x_0, y_0)$ with the smallest TDOA is the estimated event location.
2) Assuming uniform propagation speed $v$, then TDOA will be proportional to the distance between the measurement location $(x_j, y_j)$ and the event location $(x_0, y_0)$. The reference time delay, which is denoted by $\text{TDOA}'_{x,y}$, can be obtained by linear regression for each measurement using estimated event location and measured TDOAs. The linear regression slope is an indicator of the average propagation speed.
3) Measured TDOAs that are not within the 1.5 times of the inter-quartile range near $\text{TDOA}'_{x,y}$ are considered as outliers. Then outliers are deleted from measurements and the interpolation is re-performed.

Since the proposed approach is based on frequency deviation, it can non-invasively perceive system inertia distribution using synchronous measurements collected at the distribution level. The method can be directly applied to transmission-level PMU measurements.

## III. CASE STUDIES

Applying the proposed approach, Fig. 1 shows the results of the U.S. EI system. using FNET/GridEye measurements. Fig. 1 (a) used event measurements in winter 2014, while Fig. 1 (b) is based on event measurements in autumn 2014. It shows that the winter case has slower propagation speed because of larger system inertia. Comparing the two graphs, one can interpret the seasonal changes of the inertia distribution pattern.

Using simulation results of the 2030 U.S. EI dynamic model, Fig. 2 shows an example of identifying system inertia change due to variations in renewable instantaneous output and unit commitment. Fig. 2 (a) shows the propagation speed distribution for the base case with low renewable output and commitment, while Fig. 2 (b) shows the case in which PV instantaneous penetration in the New England ISO (NEISO, located in northeast EI) increased to around 45%. It shows that the proposed approach can clearly identify the inertia decrease in the New England area due to high renewable outputs.

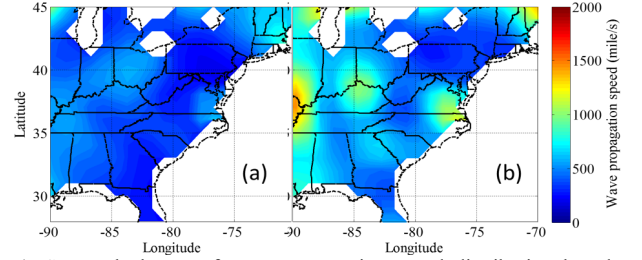

Fig. 1. Seasonal change of wave propagation speed distribution based on FNET/GridEye measurements ((a) 2014 winter and (b) 2014 autumn)

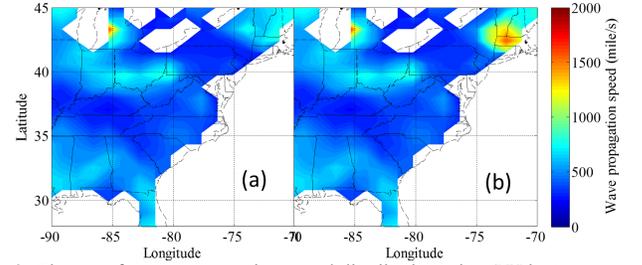

Fig. 2. Change of wave propagation speed distribution when PV instantaneous output increased in NEISO ((a) Base case; (b) PV output increases in NEISO)

## IV. CONCLUSION

This paper presented a non-invasive approach to identify the system inertia change using PMU measurements. Actual PMU data and simulated data validated the effectiveness of the approach.